\documentclass[]{spie}  

\usepackage{amsmath,amsfonts,amssymb}
\usepackage{graphicx}
\usepackage[colorlinks=true, allcolors=blue]{hyperref}
\usepackage{float}
\title{Topology-based deep-learning segmentation method for deep anterior lamellar keratoplasty (DALK) surgical guidance using M-mode OCT data}

\author[a]{Jinglun Yu}
\author[a]{Hongrui Yi}
\author[a]{Yaning Wang}
\author[b]{Justin D.~Opfermann}
\author [c,d] {William G.~Gensheimer}
\author[b]{Axel Kriger}
\author[a]{Jin U.~Kang}
\affil[a]{Department of Electrical and Computer Engineering, Johns~Hopkins~University,~Baltimore,~MD~21218,~USA}
\affil[b]{Department of Mechanical Engineering, Johns~Hopkins~University,~Baltimore,~MD~21218,~USA}
\affil[c]{Department of Ophthalmology, White~River~Junction~VA~Medical~Center,~White~River~Junction,~VT~05001,~USA.}
\affil[d]{Department of Ophthalmology, Dartmouth-Hitchcock~Medical~Center,~Lebanon,~NH~03766,~USA.}
\authorinfo{Further author information: (Send correspondence to Jinglun Yu)\\Jinglun Yu: E-mail: jyu146@jhu.edu\\}

\begin{document} 
\maketitle
\begin{abstract}
Deep~Anterior~Lamellar~Keratoplasty~(DALK) is a partial-thickness corneal transplant procedure for treating corneal stromal diseases. A key step involves the precise separation of the deep stroma from Descemet’s~membrane~(DM) using the Big~Bubble technique. To simplify the tasks of needle insertion and pneumo-dissection in this technique, we previously proposed an Optical~Coherence~Tomography~(OCT)-guided, eye-mountable robot that utilizes real-time corneal layers tracking from M-mode OCT signals for control. However, signal noise and instability during the manipulation of the OCT fiber sensor-integrated needle hinder the performance of conventional deep-learning segmentation methods, leading to rough and inaccurate detection of corneal layers. To overcome these limitations, we develop a topology-based deep-learning segmentation method that integrates a topological loss function with a modified network architecture. This approach effectively mitigates noise effects and enhances segmentation speed, precision, and stability. Validation on both \emph{in vivo}, \emph{ex vivo}, and hybrid rabbit eyes data sets demonstrates that our approach surpasses traditional loss-based techniques, delivering fast, accurate, and robust segmentation of the epithelium and DM for surgical guidance.
\end{abstract}


\section{INTRODUCTION}
\label{sec:intro} 
As a partial-thickness corneal surgery, DALK offers significant advantages over full-thickness procedures by reducing endothelial rejection rates and preserving endothelial cell density~\cite{gensheimer2024comparison}. The success of DALK, particularly when employing the Big~Bubble technique, relies heavily on accurate visualization and precise detection of corneal layers. Given the fragile nature of the deep stromal structure and the micron-scale distances between DM and endothelium, compare with the standard needle\cite{opfermann2024novel, singh2024live, xu2023neural}, the OCT fiber sensor-integrated needle serves as an ideal tool to support this procedure~\cite{wang2023common, wang2024optical}. By providing precise depth measurements and detailed layer information, this tool simplifies the surgical process, reduces complexity, and enhances procedural control.

\begin{figure} [H]
   \begin{center}
   \begin{tabular}{c} 
   \includegraphics[height=7cm]{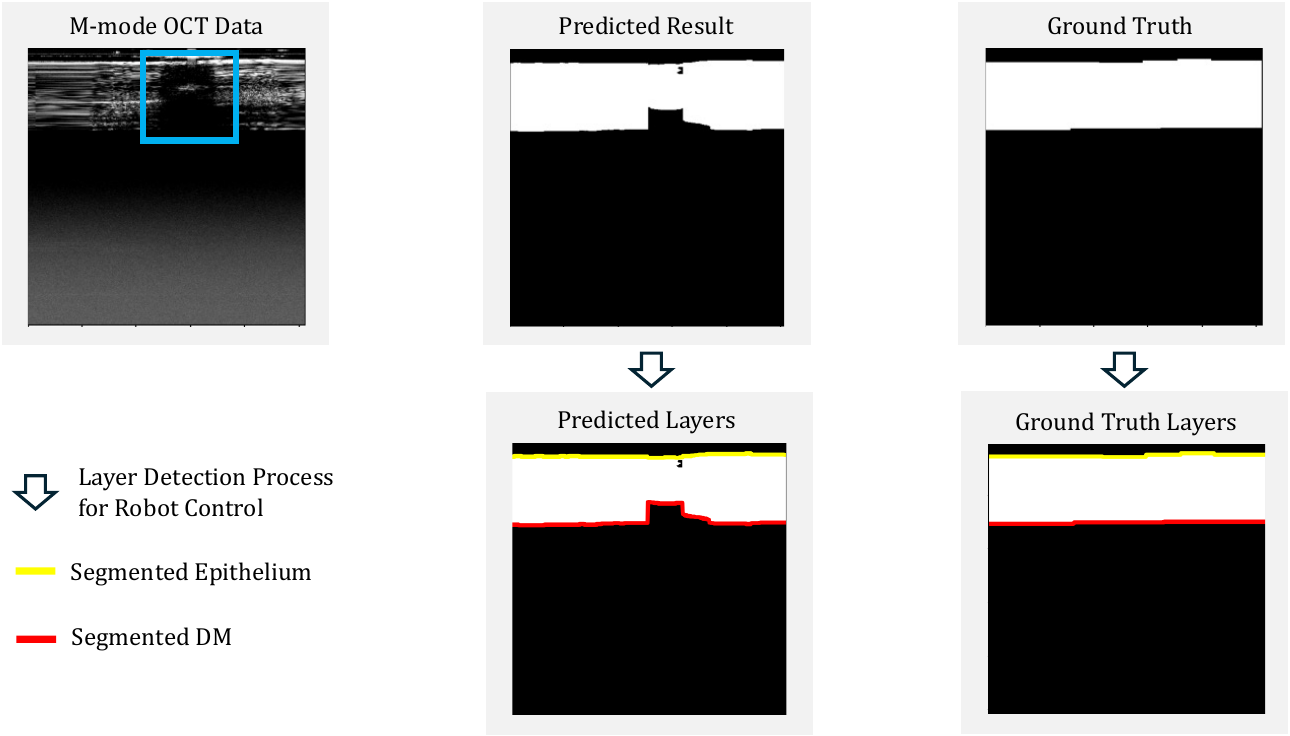}
   \end{tabular}
   \end{center}
   \caption[fig:1] 
   { \label{fig:1} Example of rough and inaccurate segmentation caused by signal noise and fluctuations. Blue Box: Hollow regions and irregular boundaries resulting from information loss and noise interference.}
   \end{figure}

To further address the technical challenges associated with needle insertion and pneumo-dissection in the Big~Bubble technique, our previous work developed an OCT-based, sensor-integrated, eye-mountable robotic system\cite{kaluna2024robotic}. Control of the robot relies on M-mode OCT signals captured by an OCT fiber sensor-integrated needle. An accurate segmentation of corneal layers from real-time OCT signals, such as the epithelium and DM, is essential for guiding the DALK procedure. However, during needle insertion and rotation, the M-mode data are inevitably affected by random noise and unstable signals. Conventional deep-learning segmentation approaches, which rely on cross-entropy or smoothness losses, struggle to compel the network to recover lost information under suboptimal data acquisition conditions \cite{wang2024reimagining}, leading to coarse and inaccurate segmentation, as illustrated in Fig.~\ref{fig:1}. Additionally, the unoptimized U-Net architecture, with the slow inference speed, reduces the real-time corneal layers tracking performance and introduces delays in surgical guidance.

In this paper, we propose a method that incorporates a topology-based loss function within a customized U-Net framework to enhance segmentation robustness and accuracy under noisy and unstable signal conditions while accelerating inference speed. Experimental results show that this approach significantly mitigates the impact of signal fluctuations and sensor motion on segmentation accuracy, enabling more precise corneal layers tracking. Furthermore, the improved processing speed enhances real-time tracking performance, providing more responsive and reliable guidance.

\section{METHODS}
\subsection{Data Collection}
\label{sec:title}
The data sets used in this paper consist of two parts. The \emph{in vivo} data set includes 500 M-mode OCT images of size $512 \times 512$, along with manually annotated labels, collected from 4 pairs of rabbit eyes in the previous robot-assisted DALK experiment\cite{wang2024reimagining}. The \emph{ex vivo} data set comprises 250 $512 \times 512$ M-mode OCT images and their corresponding labels, obtained from 12 rabbit eyes\cite{gensheimer2024comparison}. A hybrid data set is constructed by merging the \emph{in vivo} and \emph{ex vivo} data sets, providing a comprehensive collection for model training and evaluation.

\subsection{Topology-based Loss Function}
The applied hybrid loss function combines the Binary Cross Entropy (BCE) loss with topological loss\cite{mirikharaji2018star}. The hybrid loss function is defined as:
\begin{equation}
L_{hybrid} = \alpha L_{BCE} + \beta L_{T},
\end{equation}
where $\alpha$ and $\beta$ are parameters to control the contributions of the BCE loss and topological loss, in this paper $\alpha$/$\beta$ = 1/2.

The BCE loss, denoted as \( L_{BCE} \), evaluates the pixel-wise discrepancy between predicted result and ground truth (annotated label), ensuring accurate segmentation performance at the pixel level, which is defined as:
\begin{equation}
L_{BCE} = -\sum_{i=1}^{N} \left[y_i \cdot \log(\hat{y}_i) + \left(1 - y_i\right) \cdot \log(1 - \hat{y}_i) \right]
\end{equation}
where \(y_i\) represents the binary ground truth label (0 or 1) for the \( i \)-th pixel, \( \hat{y}_i \) represents the predicted probability for the \( i \)-th pixel, and \( N \) denotes the total number of pixels in the image.

\begin{figure}
   \begin{center}
   \begin{tabular}{c} 
   \includegraphics[height=5cm]{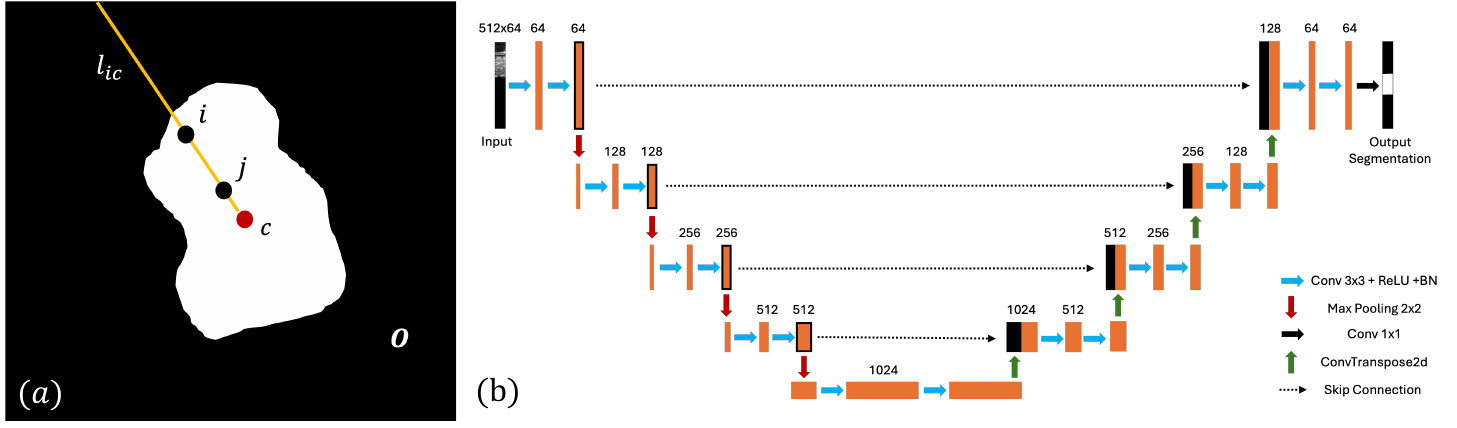}
   \end{tabular}
   \end{center}
   \caption[fig:2] 
   { \label{fig:2}\textbf{(a)}~Geometric illustration of the star shape prior in the topological loss. \textbf{(b)}~Modified U-Net architecture.}
   \vspace{0.1cm}
   \end{figure}

As shown in Fig.~\ref{fig:1}, segmentation performance is primarily affected by information loss and noise introduced during needle motions and signal fluctuations, leading to hollow areas and rough boundaries in the M-mode OCT image. To achieve more robust segmentation, we introduce the topological loss \(L_{T}\), which leverages the star shape prior\cite{mirikharaji2018star} to enforce shape and geometric constraints between segmentation regions, effectively mitigating signal artifacts and preserving essential structural details. As Fig.~\ref{fig:2}~(a) shows, assuming \( c \) is the center of a segmentation region \( O \), the star shape prior enforces that for any pixel \( i \) inside the object, all pixels \( j \) along the straight line \( l_{ic} \) connecting \( i \) to the center \( c \) must also belong to the object. The topological loss is formulated as:
\begin{equation}
L_{T} = \sum_{i=1}^N \sum_{j \in l_{ic}} B_{ij} \cdot \left| y_i - \hat{y}_i \right| \cdot \left| \hat{y}_i - \hat{y}_j \right|,
\end{equation}
where the summation over \( i \) traverses all pixels within \( O \), and \( j \in l_{ic} \) represents the pixels along the straight line segment connecting \( i \) to the region center \( c \). Here, \( B_{ij} \) is a binary indicator function that equals \( 1 \) if \( y_i = y_j \), and \( 0 \) otherwise. This term penalizes inconsistencies in predicted probabilities along \( l_{ic} \) to preserve geometric and structural continuity.

\subsection{Modified Network Structure}
In this paper, we develop a modified U-Net framework, as illustrated in Fig.~\ref{fig:2}~(b), to enhance high-quality feature extraction while achieving faster segmentation of corneal layers. Compared with our previous U-Net architecture \cite{gensheimer2024comparison}, the modified U-Net adopts a more stable and precise design. The framework integrates learnable upsampling layers with transposed convolutions and normalization techniques, facilitating accurate spatial resolution recovery and robust training dynamics. Furthermore, the use of smaller $3 \times 3$ convolution kernels significantly reduces the inference time for high-resolution image processing. This efficient feature processing and reconstruction pipeline ensures high segmentation accuracy while optimizing memory and compute utilization to accelerate inference.

\section{EXPERIMENTS AND RESULTS}
\label{sec:sections}
\subsection{Data Description}
In the experiments, the \emph{in vivo} data set is randomly split into 400 image pairs for training and 100 image pairs for testing. Similarly, the \emph{ex vivo} data set is divided into 200 image pairs for training and 50 image pairs for testing. For hybrid data experiments, the data set includes 600 image pairs for training and 150 image pairs for testing. A 20\% validation split is applied to all training data sets.

To align with the characteristics of M-mode OCT data and the demands of real-time tracking, the network automatically crops each $512 \times 512$ image pair into $8 \times 512 \times 64$ patches for input. The cropped patches are normalized using the mean and standard deviation of the corresponding training set, optimizing the performance of the proposed topology-based loss. During inference, the network reconstructs the processed patches into their original $512 \times 512$ format.
     
   \begin{figure}[H]
   \begin{center}
   \begin{tabular}{c} 
   \hspace{-0.2cm}
   \includegraphics[height=12 cm]{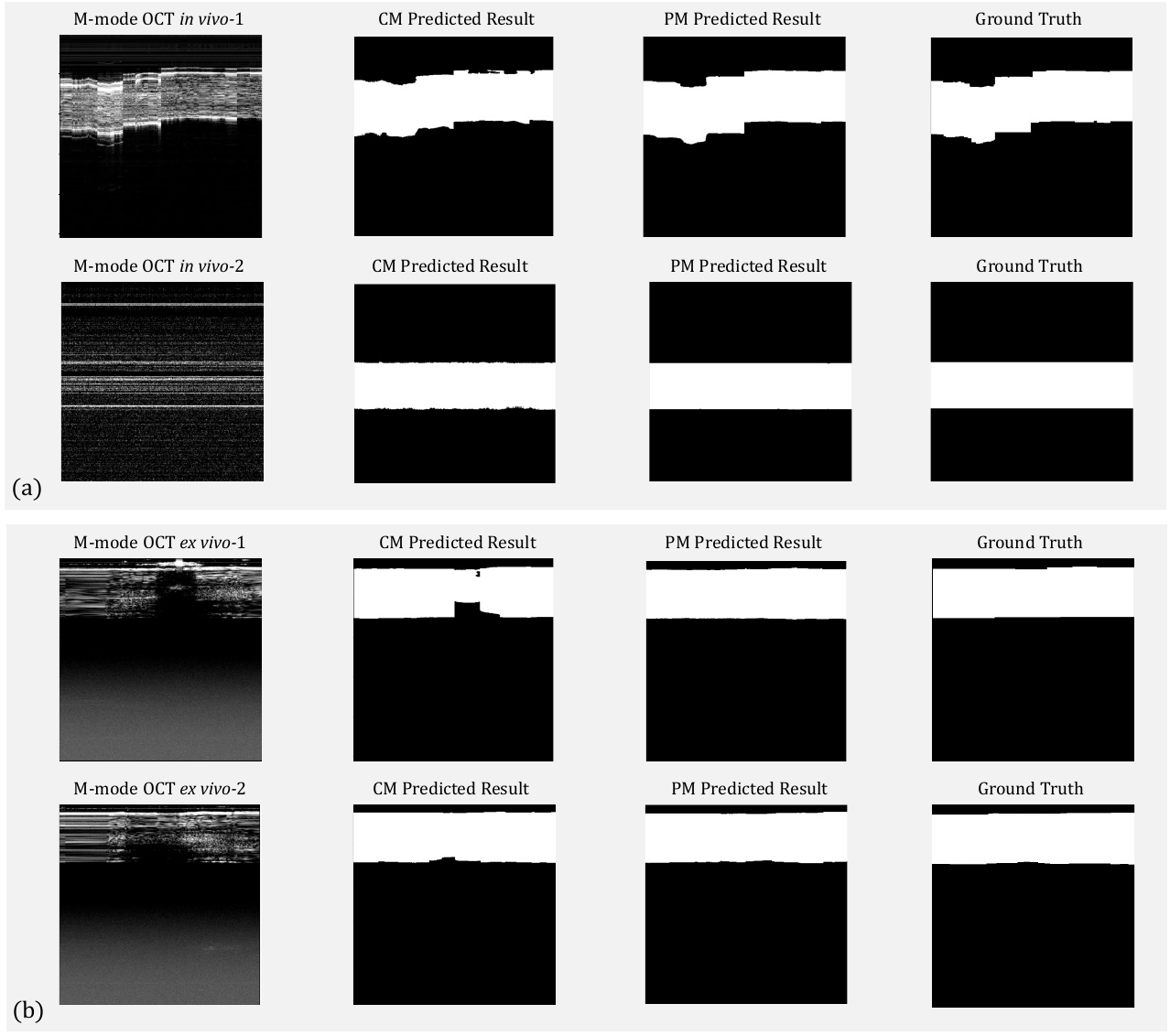}
   \end{tabular}
   \end{center}
   \caption[fig:3] 
    { \label{fig:3} Comparison of segmentation results between the proposed method (PM) and the conventional method (CM).
    \textbf{(a)}~Representative example from the \emph{in vivo} dataset.
    \textbf{(b)}~Representative example from the \emph{ex vivo} dataset. }
   \end{figure}
   
\subsection{Performance Evaluation}
The performance of the proposed method (PM) was compared to the conventional method (CM) that applied precedent U-Net architecture with conventional BCE loss\cite{gensheimer2024comparison}. All experiments were conducted on an NVIDIA GeForce RTX 4070 Ti SUPER GPU. The evaluation emphasized robustness under signal noise and fluctuation, segmentation accuracy for tracking corneal layers (epithelium and DM), and inference time efficiency to assess the potential for real-time guidance.

\begin{figure}[H]
   \begin{center}
   \begin{tabular}{c} 
   \hspace{-0.7cm}
   \vspace{-1.5cm}
   \includegraphics[height=6 cm]{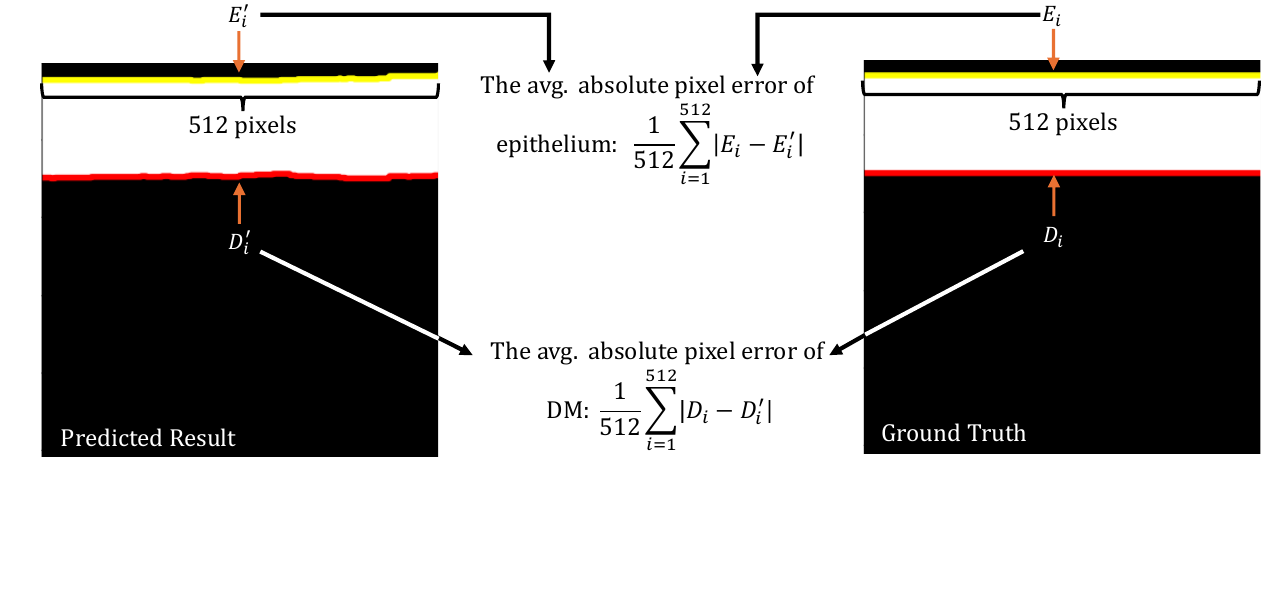}
   \end{tabular}
   \end{center}
   \caption[fig:4] 
    { \label{fig:4} 
       Evaluation pipeline for the average absolute pixel error in corneal layer segmentation. $E_i$ and $E_i^\prime$, as well as $D_i$ and $D_i^\prime$, denote the vertical pixel positions of the epithelium and Descemet's membrane for the ground truth and predicted result, respectively.
    }
  \end{figure}

Review of each method’s segmented image can specifically demonstrate the robustness under suboptimal data acquisition condition. Figure~\ref{fig:3} illustrates examples from \emph{in vivo} and \emph{ex vivo} data sets. The conventional loss-based network suffers from signal degradation, whereas the proposed method achieves accurate segmentation despite these challenges. Segmentation performance was quantified using four metrics, the average structural similarity index measure (SSIM), peak signal-to-noise ratio (PSNR), intersection over union (IoU), and Dice coefficient. These metrics were calculated for each model's prediction against the ground truth across the testing sets. Tracking performance and stability was evaluated by calculating the average absolute error of each pixel along the segmented upper (epithelium) and lower (DM) boundaries relative to the ground truth, as illustrated in Fig.~\ref{fig:4}. The average inference time across each test set was also analyzed to compare the real-time applicability of both methods.

    \begin{table}[h!]
    \caption{Performance comparison of proposed method (PM) and conventional method (CM) across different data sets.}
    \label{tab:1}
    \begin{center} 
    \begin{tabular}{|c|c|c|c|c|c|c|}
    \hline
    \textbf{Method} & \textbf{Data Set} & \textbf{SSIM $\uparrow$} & \textbf{PSNR $\uparrow$} & \textbf{IoU $\uparrow$} & \textbf{Dice $\uparrow$} & \textbf{Inference Freq. $\uparrow$} \\ \hline
    PM & \emph{in vivo}  & \textbf{0.9934} & 31.05 & \textbf{0.9910}  & \textbf{0.9955} & 35 Hz \\ \hline
    CM & \emph{in vivo}  & 0.9731 & 21.97 & 0.9542 & 0.9763 & 25 Hz \\ \hline
    PM & \emph{ex vivo}  & 0.9901 & 29.41 & 0.9874 & 0.9936 & 34 Hz \\ \hline
    CM & \emph{ex vivo}  & 0.9845 & 24.96 & 0.9798 & 0.9898 & 25 Hz \\ \hline
    PM & hybrid   & \textbf{0.9934} & \textbf{35.11} & 0.9908 & 0.9953 & \textbf{40 Hz} \\ \hline
    CM & hybrid   & 0.9858 & 25.95 & 0.9813 & 0.9905 & 28 Hz \\ \hline
    \end{tabular}
    \end{center}
    \end{table}

    \begin{table}[h!]
    \caption{Comparison of tracking accuracy between proposed method (PM) and conventional method (CM) layers across various data sets. The average absolute errors are reported in both pixels and actual distances (\(\mu\)m). For the applied M-mode data, each pixel corresponds to 2.61 \(\mu\)m.}
    \label{tab:2}
    \begin{center} 
    \begin{tabular}{|c|c|cc|cc|}
    \hline
    \textbf{Method} & \textbf{Data Set} & \multicolumn{2}{c|}{\textbf{Avg. Epithelium Error $\downarrow$}} & \multicolumn{2}{c|}{\textbf{Avg. DM Error $\downarrow$}} \\ \hline
    PM & \emph{in vivo} & 0.54 pixel& 1.4094 \(\mu\)m& \textbf{0.64} pixel& \textbf{1.6704} \(\mu\)m\\ \hline
    CM & \emph{in vivo} & 2.73 pixel& 7.1253 \(\mu\)m& 2.82 pixel& 7.3602 \(\mu\)m\\ \hline
    PM & \emph{ex vivo} & \textbf{0.34} pixel& \textbf{0.8874} \(\mu\)m& 1.26 pixel& 3.2886 \(\mu\)m\\ \hline
    CM & \emph{ex vivo} & 0.75 pixel& 1.9575 \(\mu\)m& 1.84 pixel& 4.8024 \(\mu\)m\\ \hline
    PM & hybrid & 0.38 pixel& 0.9918 \(\mu\)m& 0.67 pixel& 1.7487 \(\mu\)m\\ \hline
    CM & hybrid & 0.89 pixel& 2.3229 \(\mu\)m& 1.41 pixel& 3.6801 \(\mu\)m\\ \hline
    \end{tabular}
    \end{center}
    \end{table}

According to Tab.~\ref{tab:1}, the proposed method consistently outperformed the conventional method among all data sets in terms of average SSIM, PSNR, IoU, Dice coefficient, and inference frequency. Notably, for the \emph{in vivo} data set, our novel method achieved an average PSNR of 31.05, substantially higher than previous method's 21.97. These improvements illustrate the robustness and accuracy of the proposed method in segmenting corneal layers even under challenging \emph{in vivo} conditions. Moreover, the inference frequency of the proposed method was significantly faster across all data sets, reaching up to 40 Hz for the hybrid data set, compared to conventional method's 28 Hz, demonstrating the real-time applicability of the proposed approach. 

Tracking accuracy comparisons in Tab.~\ref{tab:2} further emphasize the advantages of our proposed segmentation method. For the \emph{ex vivo} data set, the new method reduced the average absolute error in epithelium segmentation to 0.34 pixels (0.8874 \(\mu\)m), compared to conventional-loss model's 0.75 pixels (1.9575 \(\mu\)m). For DM tracking, the new method achieved an error of 1.26 pixels (3.2886 \(\mu\)m), notably lower than conventional method’s 1.84 pixels (4.8024 \(\mu\)m). The performance gap widened further in the \emph{in vivo} data set, where the proposed method recorded epithelium and DM errors of 0.54 pixels (1.4094 \(\mu\)m) and 0.64 pixels (1.6704 \(\mu\)m), respectively, while the previous method exhibited much higher errors of 2.73 pixels (7.1253 \(\mu\)m) and 2.82 pixels (7.3602 \(\mu\)m). These differences can be attributed to the inherently complex nature of \emph{in vivo} data acquisition, where physiological factors such as eye movement, heartbeat, and respiration introduce dynamic noise and motion artifacts, making segmentation and tracking more challenging. Furthermore, \emph{ex vivo} samples often exhibit higher contrast and sharper layer boundaries due to optimized preparation and the absence of physiological interference, enabling more accurate segmentation. These results underline the proposed method's robustness and precision in tracking corneal layers under varying data acquisition conditions.

\begin{figure} [ht]
   \begin{center}
   \begin{tabular}{c} 
   \includegraphics[height=4cm]{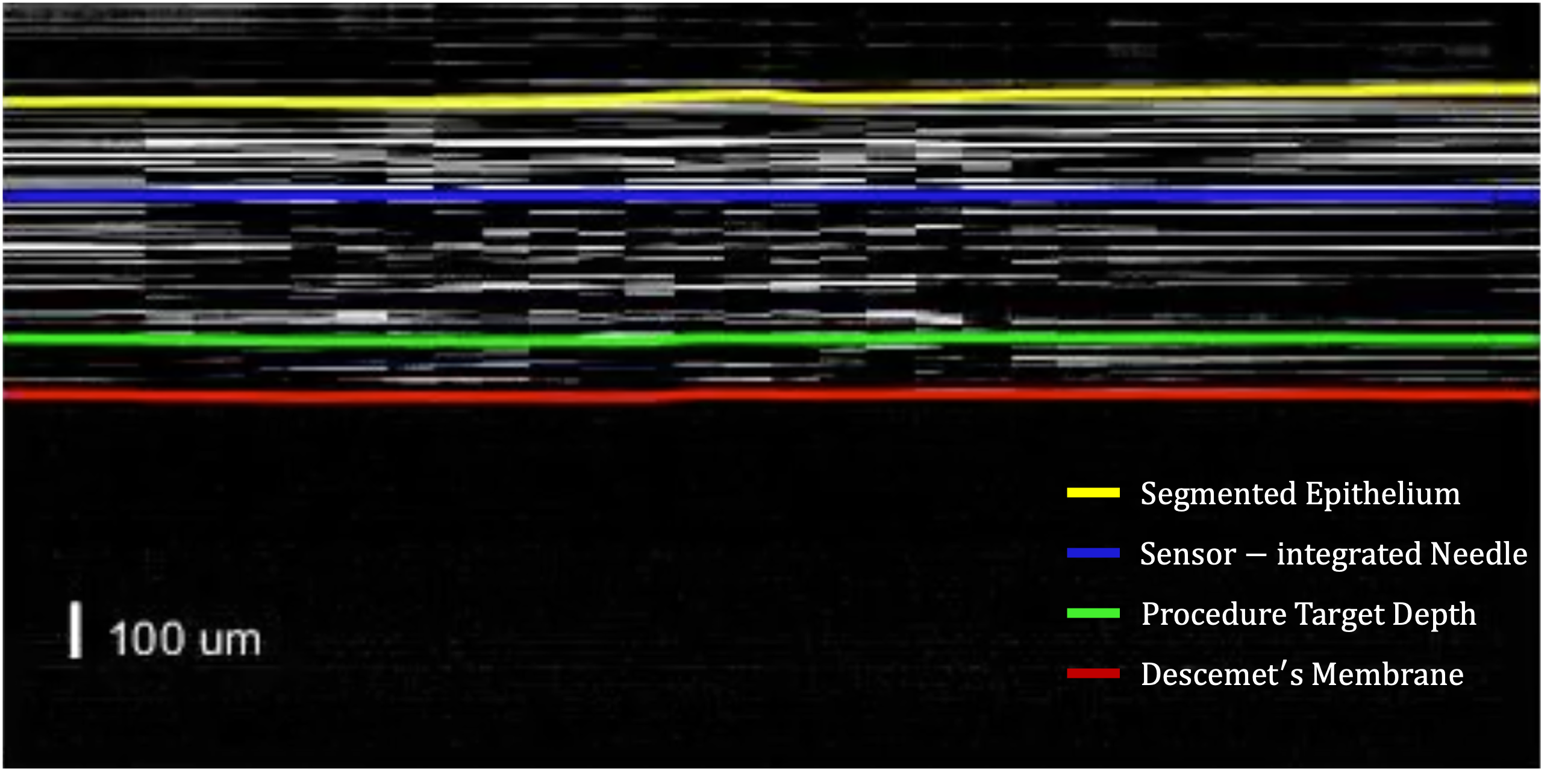}
   \end{tabular}
   \end{center}
   \caption[example] 
   { \label{fig:Video 1} Video 1-Sample video of real-time segmentation and tracking of corneal layers using the proposed method. \url{http://dx.doi.org/doi.number.goes.here}}
   \end{figure}

To demonstrate the overall improvement in corneal layer tracking for practical surgical guidance, we applied the proposed model to segment data from our OCT-based, sensor-integrated, eye-mountable robotic system\cite{kaluna2024robotic} during a real-time robotic-assisted DALK procedure. A sample video (Fig.~\ref{fig:Video 1}) showcases the process, highlighting the model's capability to reliably and seamlessly segment the epithelium and DM throughout the robotic insertion. These results underscore the potential of our approach to provide robust and dependable real-time guidance for DALK procedures.

\section{DISCUSSION AND CONCLUSION}
In this paper, we developed a deep-learning segmentation method that integrates a topology-based loss function with a modified U-Net architecture, enabling accurate segmentation of the epithelium and DM from M-mode OCT signal data. The proposed method demonstrates resilience to signal noise and fluctuations, even under challenging data acquisition conditions during surgical guidance. Experimental results show that our approach surpasses conventional loss functions and network architectures in both precision and robustness. By mitigating the effects of signal information loss during robotic surgical procedures, the method provides accurate, stable, and fast segmentation for epithelium and DM tracking, significantly enhancing real-time surgical guidance. Future work will prioritize further enhancing the robustness of the segmentation method. As the current data set is randomly split, allowing the network to potentially learn from similar samples of the same eyeball, it is essential to evaluate the method on entirely unseen eyeball data in next stage experiments.
   
\acknowledgments 
 This work was supported by National Institute of Biomedical
Imaging and Bioengineering of the National Institutes of Health Grant under award number 1R01EY032127 (PI: Jin U.~Kang). The study was conducted at Johns~Hopkins~University and Dartmouth-Hitchcock~Medical~Center.

\bibliography{report} 

\begin{thebibliography}{1}

\bibitem{gensheimer2024comparison}
Gensheimer, W.~G., Opfermann, J., Wang, Y., Kaluna, J., Krieger, A., and Kang, J.~U., ``Comparison of vertical cannula insertion techniques for big bubble deep anterior lamellar keratoplasty,'' {\em Investigative Ophthalmology \& Visual Science}~{\bf 65}(7),  3700--3700 (2024).

\bibitem{opfermann2024novel}
Opfermann, J., Wang, Y., Kaluna, J., Krieger, A., Kang, J.~U., and Gensheimer, W.~G., ``Novel vertical big bubble needle approach vs. the standard approach for deep anterior lamellar keratoplasty (dalk),'' {\em Investigative Ophthalmology \& Visual Science}~{\bf 65}(7),  3684--3684 (2024).

\bibitem{singh2024live}
Singh, M.~S., Li, K., Yu, J., Liu, X., Kang, J.~U., et~al., ``Live porcine eye model studies of subretinal injection using handheld endoscopy oct integrated injector,'' {\em Investigative Ophthalmology \& Visual Science}~{\bf 65}(7),  5499--5499 (2024).

\bibitem{xu2023neural}
Xu, J., Yu, J., Yao, J., and Zhang, R., ``The neural networks-based needle detection for medical retinal surgery,'' in [{\em International Conference on Computer Graphics, Artificial Intelligence, and Data Processing (ICCAID 2022)}{\nolinebreak\hspace{0.1em}]},   {\bf 12604},  674--678, SPIE (2023).

\bibitem{wang2023common}
Wang, Y., Guo, S., Opfermann, J.~D., Kaluna, J., Gensheimer, B.~G., Krieger, A., and Kang, J.~U., ``Common-path optical coherence tomography guided vertical pneumodissection for dalk,'' in [{\em Optical Fibers and Sensors for Medical Diagnostics, Treatment and Environmental Applications XXIII}{\nolinebreak\hspace{0.1em}]},   {\bf 12372},  15--19, SPIE (2023).

\bibitem{wang2024optical}
Wang, Y., {\em OPTICAL COHERENCE TOMOGRAPHY BASED OPHTHALMIC AND GASTROINTESTINAL SURGICAL GUIDANCE USING DEEP LEARNING}, PhD thesis, Johns Hopkins University (2024).

\bibitem{kaluna2024robotic}
Kaluna, J., Opfermann, J.~D., Wang, Y., Kang, J.~U., Gensheimer, W., and Krieger, A., ``A robotic injection system for consistent pneumo-dissection of the deep stroma in big bubble dalk surgery,'' in [{\em 2024 46th Annual International Conference of the IEEE Engineering in Medicine and Biology Society (EMBC)}{\nolinebreak\hspace{0.1em}]},   1--7, IEEE (2024).

\bibitem{wang2024reimagining}
Wang, Y., Opfermann, J., Yu, J., Yi, H., Kaluna, J., Biswas, R., Zuo, R., Gensheimer, W., Krieger, A., and Kang, J., ``Reimagining partial thickness keratoplasty: An eye mountable robot for autonomous big bubble needle insertion,'' {\em arXiv preprint arXiv:2410.14577}  (2024).

\bibitem{mirikharaji2018star}
Mirikharaji, Z. and Hamarneh, G., ``Star shape prior in fully convolutional networks for skin lesion segmentation,'' in [{\em Medical Image Computing and Computer Assisted Intervention--MICCAI 2018: 21st International Conference, Granada, Spain, September 16-20, 2018, Proceedings, Part IV 11}{\nolinebreak\hspace{0.1em}]},   737--745, Springer (2018).

\end{thebibliography}
\bibliographystyle{spiebib} 
   
\end{document}